\RequirePackage{lineno}
\documentclass[twocolumn,showpacs,superscriptaddress,amsmath,amssymb]{revtex4-2}
\usepackage{graphicx}
\usepackage{epsfig}
\usepackage{dcolumn}
\usepackage{bm}
\usepackage{overpic}
\usepackage{color}
\usepackage{ulem}
\usepackage{lineno}
\usepackage{multirow}
\usepackage{subfigure}

\begin{document}
\title{\boldmath Study of the process $e^{+}e^{-}\rightarrow\phi\eta$ at center-of-mass energies between 2.00 and 3.08~GeV}
\author{
\begin{small}
\begin{center}
M.~Ablikim$^{1}$, M.~N.~Achasov$^{10,b}$, P.~Adlarson$^{67}$, S. ~Ahmed$^{15}$, M.~Albrecht$^{4}$, R.~Aliberti$^{28}$, A.~Amoroso$^{66A,66C}$, M.~R.~An$^{32}$, Q.~An$^{63,49}$, X.~H.~Bai$^{57}$
% ...
\\
\vspace{0.2cm}
(BESIII Collaboration)\\
\vspace{0.2cm} {\it
$^{1}$ Institute of High Energy Physics, Beijing 100049, People's Republic of China\\
% ...
$^{4}$ Bochum Ruhr-University, D-44780 Bochum, Germany\\
% ...
$^{10}$ G.I. Budker Institute of Nuclear Physics SB RAS (BINP), Novosibirsk 630090, Russia\\
% ...
$^{15}$ Helmholtz Institute Mainz, Staudinger Weg 18, D-55099 Mainz, Germany\\
% ...
$^{28}$ Johannes Gutenberg University of Mainz, Johann-Joachim-Becher-Weg 45, D-55099 Mainz, Germany\\
% ...
$^{32}$ Liaoning Normal University, Dalian 116029, People's Republic of China\\
% ...
$^{49}$ State Key Laboratory of Particle Detection and Electronics, Beijing 100049, Hefei 230026, People's Republic of China\\
% ...
$^{57}$ University of Jinan, Jinan 250022, People's Republic of China\\
% ...
$^{63}$ University of Science and Technology of China, Hefei 230026, People's Republic of China\\
% ...
$^{66}$ University of Turin and INFN, (A)University of Turin, I-10125, Turin, Italy; (B)University of Eastern Piedmont, I-15121, Alessandria, Italy; (C)INFN, I-10125, Turin, Italy\\
$^{67}$ Uppsala University, Box 516, SE-75120 Uppsala, Sweden\\
% ...
$^{b}$ Also at the Novosibirsk State University, Novosibirsk, 630090, Russia\\
% ...
}\end{center}
\vskip+0.8pt
\vspace{+0.4cm}
\end{small}
}

\date{\today}
\begin{abstract}
The process $e^{+}e^{-} \rightarrow \phi\eta$ is studied at 22 center-of-mass energy points ($\sqrt{s}$) between 2.00 and 3.08 GeV using 715~pb$^{-1}$ of data collected with the BESIII detector.
The measured Born cross section of $e^{+}e^{-} \rightarrow \phi\eta$ is found to be consistent with {\textsl{BABAR}} measurements, but with improved precision. 
A resonant structure around 2.175~GeV is observed with a significance of 6.9$\sigma$ with mass ($2163.5\pm6.2\pm3.0$)~MeV/$c^{2}$ and width ($31.1_{-11.6}^{+21.1}\pm1.1$)~MeV, where the first uncertainties are statistical and the second are systematic.

\end{abstract}

\maketitle

\section{\boldmath Introduction}

The first observation of the $\phi(2170)$ meson was reported by the {\textsl{BABAR}} collaboration in the initial-state-radiation (ISR) process $e^{+}e^{-}\rightarrow\gamma_{ISR}\phi\pi^{+}\pi^{-}$~\cite{Y2175BABAR1}. 
The {\textsl{BABAR}}~\cite{Y2175BABAR2,Y2175BABAR3,Y2175BABAR5}, BES~\cite{Y2175BES}, Belle~\cite{Y2175BELLE} and BESIII~\cite{Y2175BESIII,Dong,Yankun,KKpipi,Yankun2,Dong2} collaborations also studied $\phi(2170)$ decays.
Since the $\phi(2170)$ is produced in $e^{+}e^{-}$ collisions, its quantum numbers are $J^{PC}~=~1^{--}$. 
The discovery of the $\phi(2170)$ has sparked extensive discussions about its internal structure.
Proposed explanations include: a $s\bar{s}g$ hybrid state~\cite{Y2175hybrid1,Y2175hybrid2,Y2175hybrid3,Y2175hybrid5}, either the $2^{3}D_{1}$~\cite{Y2175ss2D1,Y2175ss2D2,Y2175QLi}  or $3^{3}S_{1}$~\cite{Y2175ss3S1,Y2175ss3S2} $s\bar{s}$ state, a $s\bar{s}s\bar{s}$ tetraquark state~\cite{Y2175tetraquark1,Y2175tetraquark2,Y2175tetraquark3,Y2175tetraquark4,Y2175tetraquark5,Y2175tetraquark6,Y2175tetraquark7,Y2175tetraquark8,Y2175tetraquark9}, a $\Lambda\bar{\Lambda}$ molecular state~\cite{Y2175lambda1,Y2175lambda2,Y2175lambda3,Y2175lambda4,Y2175lambda5,Y2175lambda6}, a $\phi f_{0}(980)$ resonance~\cite{Y2175tetraquark7,Y2175ordinary2,Y2175ordinary3,Y2175ordinary5} including final state interaction effects~\cite{Y2175BESIII,Y2175Swave}, a $S$-wave threshold effect~\cite{Y2175Swave}, and a $X(2240)$ state~\cite{Y2175X2240}.
None of these explanations has been able to describe all experimental observations.
Therefore, the nature of the $\phi(2170)$ still needs to be clarified by further theoretical and experimental efforts.

For different hypotheses regarding the internal structure of the $\phi(2170)$, certain decay modes such as $\phi(2170)\rightarrow\phi\eta$
~\cite{Y2175BABAR3,Y2175hybrid2,Y2175ss2D1,Y2175ss2D2} can have decay rates that vary strongly across those models.
According to the Okubo-Zweig-Iizuka (OZI) rule~\cite{OZI_rule1,OZI_rule2,OZI_rule3}, excited $\phi$ mesons are predicted to decay with a considerable fraction into the $s\bar{s}$ modes $\phi\eta$ and $\phi\eta'$~\cite{Y2175ss3S1,Y2175ss3S2}.
Both the $2^{3}D_{1}$~\cite{Y2175ss2D1,Y2175ss2D2} and $3^{3}S_{1}$~\cite{Y2175ss3S1,Y2175ss3S2} $s\bar{s}$ states as well as $s\bar{s}s\bar{s}$~\cite{Y2175tetraquark1,Y2175tetraquark2,Y2175tetraquark3,Y2175tetraquark4,Y2175tetraquark5,Y2175tetraquark6,Y2175tetraquark7,Y2175tetraquark8,Y2175tetraquark9} tetraquark states favor the $\phi\eta$ and $\phi\eta'$ decay modes according to model calculations.
Consequently, $e^{+}e^{-} \rightarrow\phi\eta$ is an excellent channel to study excited $\phi$ states~\cite{OZI_rule1,OZI_rule2,OZI_rule3}.

The {\textsl{BABAR}} collaboration observed evidence of $\phi(2170)$ in the measurement of the cross section of $e^{+}e^{-}\rightarrow\gamma_{ISR}\phi\eta$ using the ISR method~\cite{Y2175BABAR3}, and also found a hint of the $\phi(2170)$ in the process $e^{+}e^{-}\rightarrow\gamma_{ISR}\phi\eta'$~\cite{Y2175phietap}. 
Assuming that the observed structure in the process $e^{+}e^{-} \rightarrow \phi\eta$ is the $\phi(2170)$, {\textsl{BABAR}} measured a partial width of $\mathcal{B}_{\phi\eta}^{\phi(2170)}\Gamma_{e^{+}e^{-}}^{\phi(2170)}=(1.7\pm0.7\pm1.3)$~eV~\cite{Y2175BABAR3}.
In recent studies of the processes $e^{+}e^{-}\rightarrow K^{+}K^{-}$~\cite{Dong,KK2}, $\phi K^{+}K^{-}$~\cite{Yankun}, $K^{+}K^{-}\pi^{0}\pi^{0}$~\cite{KKpipi}, $\omega\eta$~\cite{Dong2}, and $\phi\eta'$~\cite{Yankun2} by the BESIII collaboration, a clear structure around 2.2~GeV is observed in the line shape of the measured cross sections.

The ratio between the $\phi\eta$ and $\phi\eta'$ partial widths is an
important observable to assess $\phi(2170)$ as a hybrid state.  An
$s\bar{s}g$ hybrid state is expected to have a stronger coupling to
$\phi\eta$, with the partial width expected to be larger than that to
$\phi\eta'$ by factors ranging from 3 up to
200~\cite{Y2175hybrid1,Y2175hybrid2}.  
The resulting partial width in the study of $e^{+}e^{-}\rightarrow\phi\eta'$ at BESIII is found to be $\mathcal{B}_{\phi\eta'}^{\phi(2170)}\Gamma_{e^{+}e^{-}}^{\phi(2170)}=(7.1\pm0.7\pm0.7)$~eV~\cite{Yankun2}.
The precision of the ratio of partial widths between the decays to $\phi\eta$ and $\phi\eta'$ of $\mathcal{B}^{\phi(2170)}\Gamma_{e^{+}e^{-}}^{\phi(2170)}$ provides a strong benchmark for theoretical models aiming to explain the nature of the $\phi(2170)$.
Hence, a new measurement of the cross section of the process $e^{+}e^{-} \rightarrow\phi\eta$ is important in order to improve our understanding of the nature of the $\phi(2170)$ resonance.

In this paper, we present an improved measurement of the Born cross section ($\sigma_{\phi\eta}^{{\rm{Born}}}$) of the process $e^{+}e^{-}\rightarrow\phi\eta$ at 22 ($\sqrt{s}$) in the range between 2.00 and 3.08~GeV with a data sample corresponding to an integrated luminosity of 715~pb$^{-1}$ collected with the BESIII experiment.

\section{\boldmath BESIII experiment and Monte Carlo simulation}

The BESIII detector is a magnetic spectrometer~\cite{BESIII} located at the Beijing Electron Positron Collider (BEPCII)~\cite{Yu:IPAC2016-TUYA01}. 
The cylindrical core of the BESIII detector consists of a helium-based multilayer drift chamber (MDC), a plastic scintillator time-of-flight system (TOF), and a CsI(Tl) electromagnetic calorimeter (EMC), which are all enclosed in a superconducting solenoidal magnet providing a 1.0~T magnetic field. 
The solenoid is supported by an octagonal flux-return yoke instrumented with resistive plate counters interleaved with steel, which serve as muon identifiers. 
The acceptance of charged particles and photons is 93\% of the full solid angle. 
The charged-particle momentum resolution at $1~{\rm GeV}/c$ is $0.5\%$, and the $dE/dx$ resolution is $6\%$ for electrons from Bhabha scattering. 
The EMC measures photon energies with a resolution of $2.5\%$ ($5\%$) at $1$~GeV in the barrel (end cap) region. 
The time resolution of the TOF barrel part is 68~ps, while that of the end cap part is 110~ps. 

The detector response, including the interaction of secondary particles with the detector material, is simulated using
{\sc geant4}~\cite{geant4} based Monte Carlo (MC) software. MC simulation
samples of 2.5 million $e^{+}e^{-}\rightarrow\phi\eta$ events per
energy point generated by using $P$-wave in the production process with the {\sc conexc}~\cite{Lei1} generator are used
for the efficiency determination and to calculate the correction factors for
radiative effects up to next-to-leading order (NLO), as well for the effect of vacuum polarization (VP).
MC samples of inclusive hadronic events generated with {\sc conexc} combined with {\sc luarlw}~\cite{Lei1} are used for background studies. 

\section{\boldmath Event selection and background analysis}
To select $e^{+}e^{-}\rightarrow\phi\eta$, the $\phi$ and $\eta$
candidates are reconstructed through their decays to $K^{+}K^{-}$ and
$\gamma\gamma$, respectively.  Selected events must have exactly two
charged tracks with opposite charge.  Tracks are reconstructed using
the MDC, and all track candidates have to be within the MDC acceptance
of $|\cos\theta|<0.93$, where $\theta$ is the polar angle with respect
to the symmetry axis of the drift chamber.
Additionally, both tracks are required to have their point of closest
approach to the interaction point be within 10~cm along the beam
direction and 1~cm in the transverse plane.  The TOF and the $dE/dx$
information are combined to calculate a particle identification (PID)
likelihood for the $\pi$, $K$, and $p$ hypotheses.  For both tracks,
it is required that the likelihood of a kaon assignment is larger than
the two alternative hypotheses.

Photon candidates are selected from showers in the EMC that are not associated with charged tracks. 
Good photon candidates reconstructed in the barrel part of the EMC must have a polar angle within $|\cos\theta|<0.8$, while photon candidates reconstructed in the end caps must have a polar angle within $0.86<|\cos\theta|<0.92$.
In order to suppress the background from ISR processes, the energy of
all photon candidates is required to be larger than 70~MeV. 
To suppress electronic noise and energy deposits unrelated to the event, the timing information from the EMC is required to be within 700~ns of the event start time for all photon candidates.

\begin{figure}[htbp]
\begin{center}
\begin{overpic}[width=8.5cm,height=6cm,angle=0]{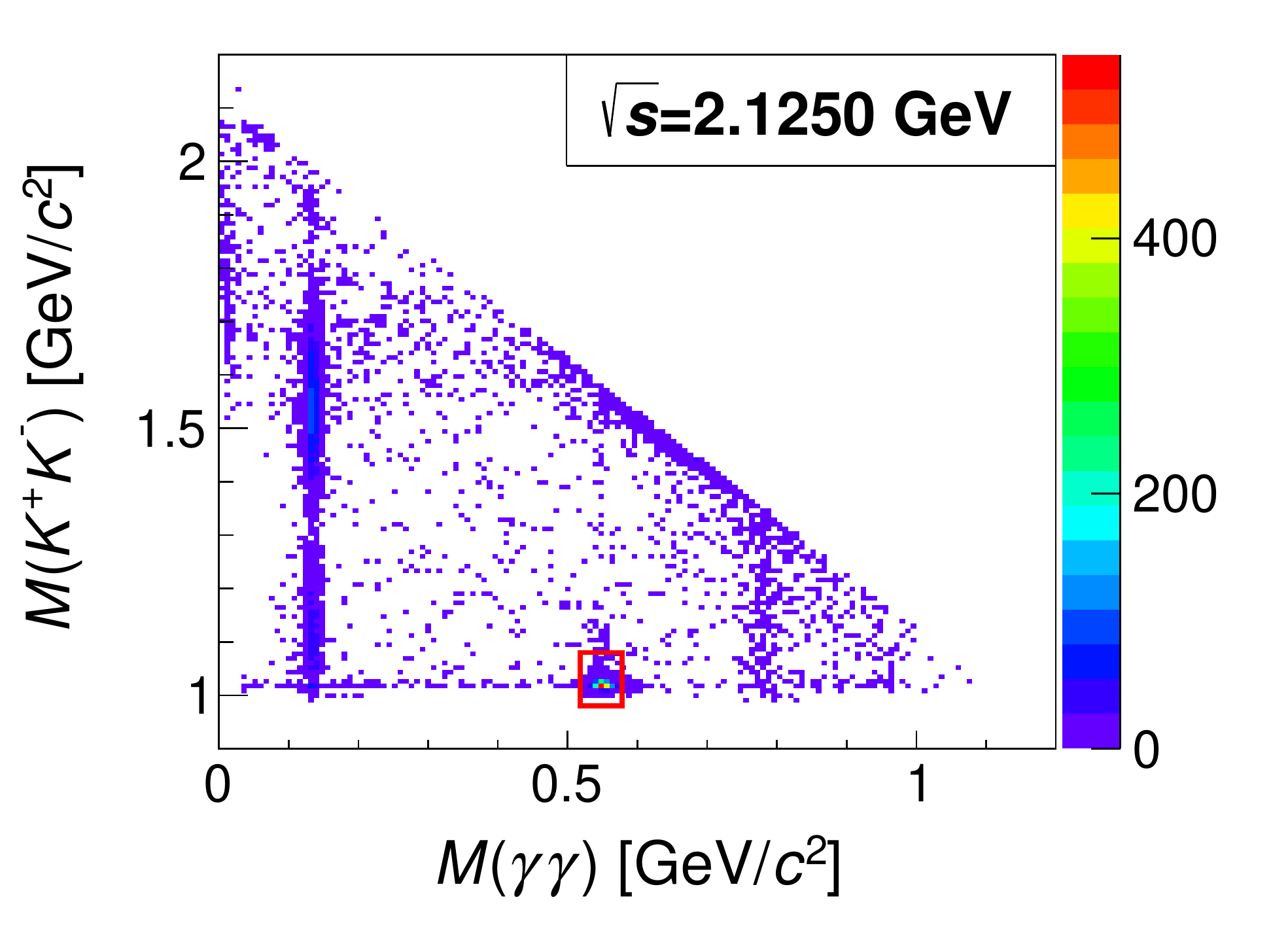}
\end{overpic}
\caption{Distribution of $m(K^{+}K^{-})$ versus $m(\gamma\gamma)$ at $\sqrt{s}=2.125$~GeV. The red rectangle indicates the signal region.
}
\label{fig::masswindow}
\end{center}
\end{figure}

A four-constraint (4C) kinematic fit is applied using the hypothesis
$e^{+}e^{-}\rightarrow K^{+}K^{-}\gamma\gamma$, constraining the
measured four-momenta of all particles to the initial center-of-mass (c.m.)
four-momentum.  For events with more than two good photon candidates,
the combination with the smallest $\chi^{2}$ of the kinematic fit is
retained for further study.  Only events with
$\chi_{4C}^{2}(K^{+}K^{-}\gamma\gamma)<100$ are kept.  In order to
suppress background contributions from the reaction $e^{+}e^{-}
\rightarrow\gamma_{ISR}\phi$, $\phi\rightarrow K^{+}K^{-}$, where
$\gamma_{ISR}$ is the detected ISR photon, a second
kinematic fit is used testing the $e^{+}e^{-}\rightarrow K^{+}K^{-}\gamma$ hypothesis. Events are
rejected if the $\chi^{2}$ of the kinematic fit to the $e^{+}e^{-}\rightarrow K^{+}K^{-}\gamma$
hypothesis is smaller than the one for the signal hypothesis.
%
%
The distribution of the invariant mass of the two kaons ($M(K^{+}K^{-})$) versus the invariant mass of the two photons ($M(\gamma\gamma)$) is shown in Fig.~\ref{fig::masswindow} for the data at a ($\sqrt{s}$) of 2.125~GeV using the above selection criteria.
An enhancement at the $\eta$ and $\phi$ meson masses is observed.
Candidate events of $e^{+}e^{-} \rightarrow \phi\eta$ are required to be within the combined $\eta$ and $\phi$ signal region, defined as $|M(\gamma\gamma)-m_{\eta}| <30$~MeV/$c^{2}$, $0.98<M(K^{+}K^{-})<1.08$~GeV/$c^{2}$, where $m_{\eta}$ is the mass of $\eta$ meson as listed by the PDG~\cite{PDG}, and 30~MeV/$c^{2}$ corresponds to three times the detector resolution at the $\eta$ mass.

The main sources of the remaining background are processes of the type $e^+e^-\to K^{+}K^{-}\eta$ (where the $K^+K^-$ pair does not originate from a $\phi$ decay), $e^+e^-\to K^{+}K^{-}\pi^{0}$ and $e^+e^-\to K^{+}K^{-}\pi^{0}\pi^{0}$.
Possible contaminations are estimated to be less than 1.0\% from studies performed on inclusive hadronic MC samples.
After selecting $\eta$ candidates using the invariant mass $M(\gamma\gamma)$, background contributions are highly suppressed and do not exhibit narrow structures in $M(K^+K^-)$ so that they can be described by a smooth polynomial function.
%
%


\section{\boldmath Cross Sections of $e^{+}e^{-}\rightarrow\phi\eta$}
\subsection{\boldmath Signal yields}
\begin{figure}[htbp]
\begin{center}
\begin{overpic}[width=8.5cm,height=6cm,angle=0]{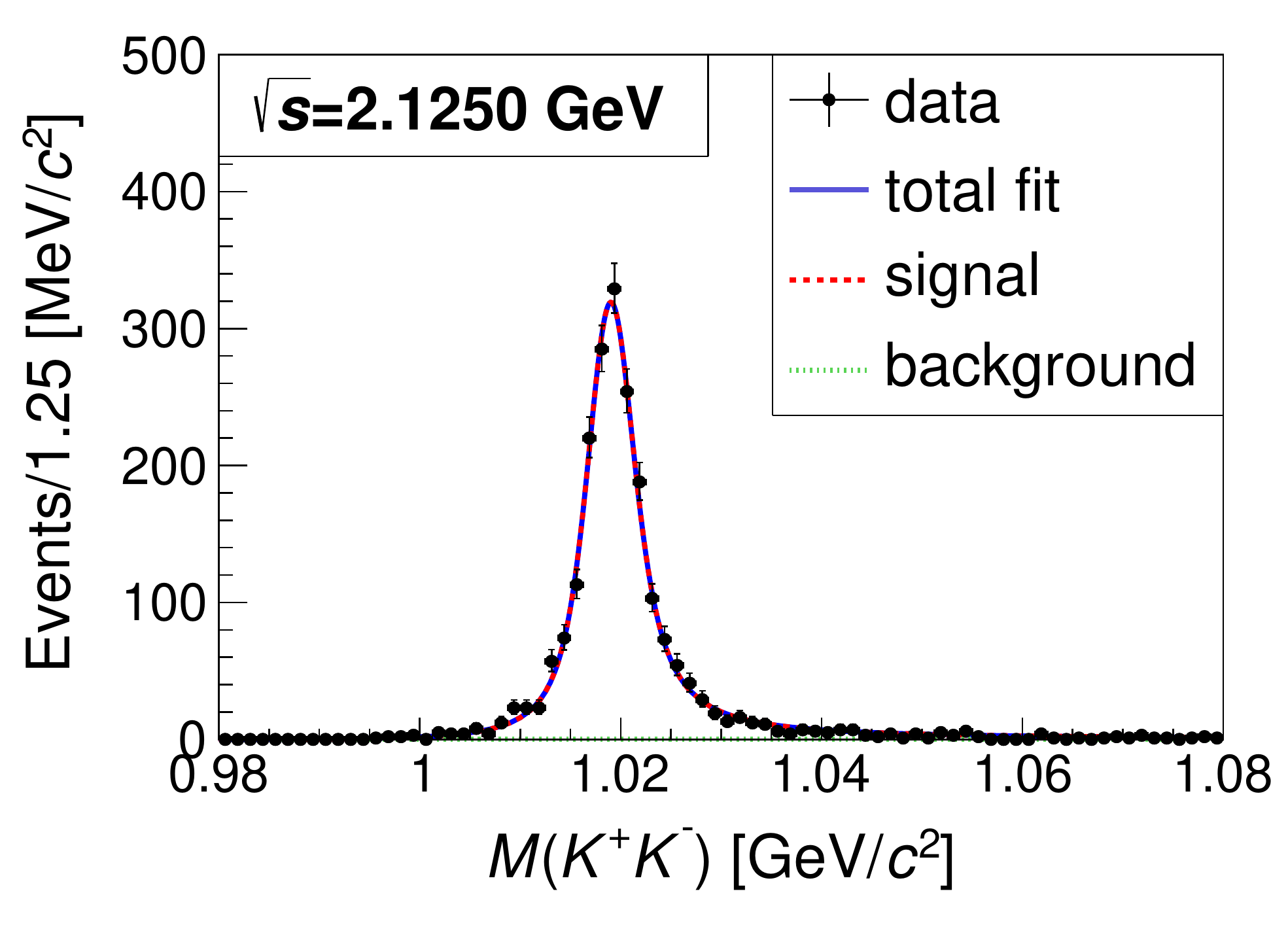}
\end{overpic}
\caption{The invariant mass of the $K^{+}K^{-}$ pair after selection of the $\eta$ meson candidate for the $\sqrt{s}$ of 2.125~GeV. Black dots represent experimental data, the blue solid curve is the total fit result, while the red dashed and green dotted curves show the individual signal and background components, respectively.}
\label{fig::fitting_CenterValue}
\end{center}
\end{figure}

The number of events of the process $e^{+}e^{-}\rightarrow\phi\eta$ is
determined by an unbinned maximum likelihood fit to the $K^{+}K^{-}$
invariant mass with a signal shape, which is parameterized by a $P$-wave
relativistic Breit-Wigner function convolved with a Gaussian function.
The background shape is described by a first order polynomial.
According to Ref.~\cite{formula_p-wave}, the $P$-wave relativistic
Breit-Wigner amplitude for $\phi\rightarrow K^{+}K^{-}$ is
\begin{eqnarray}
\label{eq:p-wave1}
\begin{split}
    A[M(K^+K^-)]=&p_{K}\Big\{M(K^+K^-)^{2}-m^{2}_{\phi}+\mathit{i}M(K^+K^-)\\
    &\cdot\Gamma[M(K^+K^-)]\Big\}^{-1}\cdot\frac{B(p_{K})}{B(p'_{K})},
\end{split}
\end{eqnarray}
where $m_{\phi}$ is fixed to the mass of the $\phi$ meson~\cite{PDG}, 
$p_{K}$ is the kaon momentum in the $\phi$ rest frame and $p'_{K}$ is the kaon momentum at the nominal $\phi$ mass. 
The width $\Gamma[M(K^+K^-)]$ is given by
\begin{eqnarray}
\label{eq:p-wave2}
 \Gamma[M(K^+K^-)]=\Bigl(\frac{p_{K}}{p'_{K}}\Bigr)^{3}\Bigl[\frac{M(K^+K^-)}{m_{\phi}}\Bigr]\Gamma_{0}\Bigl[\frac{B(p_{K})}{B(p'_{K})}\Bigr],
\end{eqnarray}
where $\Gamma_0$ is fixed to the nominal width of the $\phi$
meson~\cite{PDG} and $B(p)$ is the $P$-wave Blatt-Weisskopf form factor
with $B(p)=\sqrt{\frac{2(Rp)^2}{1+(Rp)^2}}$, where $R=3~\textrm{GeV}^{-1}c$ is chosen as the radius in the calculation of the centrifugal barrier factor.
The amplitude squared convolved with the Gaussian function is added
incoherently to the background polynomial.  The parameters of the
polynomial and the Gaussian function are free in the fit. The latter
is used to compensate for absolute resolution and an offset
of the mass scaling in data.  The fit result for the data at
$\sqrt{s}=2.125$~GeV is shown in Fig.~\ref{fig::fitting_CenterValue},
while the signal event yields $N_{{\rm{Signal}}}$ for all energy
points are summarized in Table~\ref{table_XS_result}.

\subsection{\boldmath Efficiency and radiative corrections}
\label{subsec::effisr}
With $N_{{\rm{Signal}}}$ determined, the Born cross section $\sigma_{\phi\eta}^{{\rm{Born}}}(s)$ of the process $e^{+}e^{-}\rightarrow\phi\eta$ at the c.m.~energy squared $s$ can be determined using
\begin{eqnarray}
\label{eq_sigma}
    \sigma_{\phi\eta}^{{\rm{Born}}}(s)=\frac{N_{{\rm{Signal}}}}{\mathcal{L}\cdot\epsilon\cdot(1+\delta)\cdot\frac{1}{(1-\Pi)^{2}}\cdot\mathcal{B}},
\end{eqnarray}
where $\mathcal{L}$ is the integrated luminosity measured with large angle Bhabha scattering~\cite{cite_lumi}, $\epsilon$ is the reconstruction efficiency, $(1+\delta)$ is the radiative correction factor and $\frac{1}{(1-\Pi)^{2}}$ is the VP correction factor. The explicit ($\sqrt{s}$) dependence of those variables is omitted here.
The total branching fraction $\mathcal{B}$ is the product of the branching fractions for the decays contained in the full decay chain $\mathcal{B}=\mathcal{B}(\phi\rightarrow K^{+}K^{+})\cdot\mathcal{B}(\eta\rightarrow\gamma\gamma)=(19.39\pm0.22)$\%~\cite{PDG}.
The product $\epsilon(1+\delta)$ is determined in an iterative procedure. 
At each step of the iteration, a set of 1000 MC samples is produced taking into account the fit to the Born cross section introduced in Sec.~\ref{lineshape}.
The MC samples are produced by sampling the model parameters according to a Gaussian distribution whose width is set equal to the uncertainties of the model parameters as obtained in the fit.
Each of the 1000 MC samples gives a new value of $\epsilon(1+\delta)$. 
The mean of those 1000 values is used to recalculate the Born cross section. This process is repeated until the resulting Born cross section converges. After two iterations, the observed change in the Born cross section $\sigma_{\phi\eta}^{{\rm{Born}}}(s)$ is smaller than the uncertainty of the generator, which is 0.5\%.
The efficiency, the radiative, VP correction factors and the results for $\sigma_{\phi\eta}^{{\rm{Born}}}$ are summarized in Table~\ref{table_XS_result}.

\begin{table*} [htbp]
  \begin{center}
  \footnotesize
  \caption{The integrated luminosity $\mathcal{L}$,
    the number of $\phi\eta$ events  $N_{{\rm{Signal}}}$, efficiency $\epsilon$, radiative correction factor $(1+\delta)$, VP correction factor $\frac{1}{(1-\Pi)^{2}}$ and the Born cross section $\sigma_{\phi\eta}^{{\rm{Born}}}$ for the different $\sqrt{s}$. The statistical uncertainty on the efficiency is negligible.}
  \resizebox{!}{4.7cm}{\begin{tabular}{c c c c c c c}
      \hline \hline
      $\sqrt{s}$[GeV] & $\mathcal{L}[$pb$^{-1}]$ & $N_{{\rm{Signal}}}$ &$\epsilon[\%]$ &$(1+\delta)$&$\frac{1}{(1-\Pi)^{2}}$ & $\sigma_{\phi\eta}^{{\rm{Born}}}$[pb]\\ \hline 
~~~~~$2.0000$~~~~~&~~~~~$10.1\pm0.1$~~~~~&~~~~~$237.0\pm15.4$~~~~~&~~~~~$21.5$~~~~~&~~~~~$1.113$~~~~~&~~~~~$1.037$~~~~~&~~~~~$489.9\pm31.8\pm20.0$~~~~~\\
~~~~~$2.0500$~~~~~&~~~~~$3.34\pm0.03$~~~~~&~~~~~$72.1\pm8.5$~~~~~&~~~~~$22.4$~~~~~&~~~~~$1.131$~~~~~&~~~~~$1.038$~~~~~&~~~~~$422.4\pm49.8\pm19.0$~~~~~\\
~~~~~$2.1000$~~~~~&~~~~~$12.2\pm0.1$~~~~~&~~~~~$252.0\pm15.9$~~~~~&~~~~~$23.3$~~~~~&~~~~~$1.135$~~~~~&~~~~~$1.039$~~~~~&~~~~~$389.2\pm24.5\pm16.6$~~~~~\\
~~~~~$2.1250$~~~~~&~~~~~$108\pm1$~~~~~&~~~~~$2097.0\pm45.8$~~~~~&~~~~~$23.9$~~~~~&~~~~~$1.120$~~~~~&~~~~~$1.039$~~~~~&~~~~~$358.6\pm7.8\pm14.0$~~~~~\\
~~~~~$2.1500$~~~~~&~~~~~$2.84\pm0.02$~~~~~&~~~~~$60.8\pm7.8$~~~~~&~~~~~$24.7$~~~~~&~~~~~$1.090$~~~~~&~~~~~$1.040$~~~~~&~~~~~$394.1\pm50.5\pm16.7$~~~~~\\
~~~~~$2.1750$~~~~~&~~~~~$10.6\pm0.1$~~~~~&~~~~~$146.7\pm12.6$~~~~~&~~~~~$24.4$~~~~~&~~~~~$1.348$~~~~~&~~~~~$1.040$~~~~~&~~~~~$208.4\pm17.9\pm11.0$~~~~~\\
~~~~~$2.2000$~~~~~&~~~~~$13.7\pm0.1$~~~~~&~~~~~$161.0\pm12.7$~~~~~&~~~~~$21.6$~~~~~&~~~~~$1.358$~~~~~&~~~~~$1.040$~~~~~&~~~~~$198.9\pm15.7\pm8.2$~~~~~\\
~~~~~$2.2324$~~~~~&~~~~~$14.5\pm0.1$~~~~~&~~~~~$125.9\pm11.9$~~~~~&~~~~~$22.4$~~~~~&~~~~~$1.315$~~~~~&~~~~~$1.041$~~~~~&~~~~~$145.9\pm13.8\pm7.3$~~~~~\\
~~~~~$2.3094$~~~~~&~~~~~$21.1\pm0.1$~~~~~&~~~~~$201.1\pm14.2$~~~~~&~~~~~$23.0$~~~~~&~~~~~$1.314$~~~~~&~~~~~$1.041$~~~~~&~~~~~$156.0\pm11.0\pm6.6$~~~~~\\
~~~~~$2.3864$~~~~~&~~~~~$22.5\pm0.2$~~~~~&~~~~~$197.5\pm15.1$~~~~~&~~~~~$23.3$~~~~~&~~~~~$1.342$~~~~~&~~~~~$1.041$~~~~~&~~~~~$138.8\pm10.6\pm5.9$~~~~~\\
~~~~~$2.3960$~~~~~&~~~~~$66.9\pm0.5$~~~~~&~~~~~$502.8\pm23.4$~~~~~&~~~~~$23.4$~~~~~&~~~~~$1.346$~~~~~&~~~~~$1.041$~~~~~&~~~~~$118.1\pm5.5\pm5.1$~~~~~\\
~~~~~$2.4000$~~~~~&~~~~~$3.42\pm0.03$~~~~~&~~~~~$26.3\pm5.3$~~~~~&~~~~~$23.9$~~~~~&~~~~~$1.348$~~~~~&~~~~~$1.041$~~~~~&~~~~~$118.6\pm24.0\pm6.1$~~~~~\\
~~~~~$2.6444$~~~~~&~~~~~$33.7\pm0.2$~~~~~&~~~~~$156.0\pm12.5$~~~~~&~~~~~$23.1$~~~~~&~~~~~$1.475$~~~~~&~~~~~$1.039$~~~~~&~~~~~$67.3\pm5.4\pm2.6$~~~~~\\
~~~~~$2.6464$~~~~~&~~~~~$34.0\pm0.3$~~~~~&~~~~~$157.0\pm12.5$~~~~~&~~~~~$23.0$~~~~~&~~~~~$1.476$~~~~~&~~~~~$1.039$~~~~~&~~~~~$67.5\pm5.4\pm2.8$~~~~~\\
~~~~~$2.9000$~~~~~&~~~~~$105\pm1$~~~~~&~~~~~$262.8\pm16.2$~~~~~&~~~~~$21.9$~~~~~&~~~~~$1.639$~~~~~&~~~~~$1.033$~~~~~&~~~~~$34.8\pm2.1\pm1.7$~~~~~\\
~~~~~$2.9500$~~~~~&~~~~~$15.9\pm0.1$~~~~~&~~~~~$29.0\pm5.4$~~~~~&~~~~~$21.4$~~~~~&~~~~~$1.677$~~~~~&~~~~~$1.029$~~~~~&~~~~~$25.5\pm4.7\pm1.1$~~~~~\\
~~~~~$2.9810$~~~~~&~~~~~$16.1\pm0.1$~~~~~&~~~~~$29.9\pm5.7$~~~~~&~~~~~$21.3$~~~~~&~~~~~$1.703$~~~~~&~~~~~$1.025$~~~~~&~~~~~$25.8\pm4.9\pm1.4$~~~~~\\
~~~~~$3.0000$~~~~~&~~~~~$15.9\pm0.1$~~~~~&~~~~~$29.1\pm5.6$~~~~~&~~~~~$21.1$~~~~~&~~~~~$1.720$~~~~~&~~~~~$1.021$~~~~~&~~~~~$25.6\pm4.9\pm1.8$~~~~~\\
~~~~~$3.0200$~~~~~&~~~~~$17.3\pm0.1$~~~~~&~~~~~$25.0\pm5.0$~~~~~&~~~~~$20.9$~~~~~&~~~~~$1.741$~~~~~&~~~~~$1.014$~~~~~&~~~~~$20.2\pm4.0\pm1.7$~~~~~\\
~~~~~$3.0500$~~~~~&~~~~~$14.9\pm0.1$~~~~~&~~~~~$21.0\pm4.6$~~~~~&~~~~~$20.8$~~~~~&~~~~~$1.782$~~~~~&~~~~~$0.996$~~~~~&~~~~~$19.7\pm4.3\pm0.8$~~~~~\\
~~~~~$3.0600$~~~~~&~~~~~$15.0\pm0.1$~~~~~&~~~~~$29.1\pm5.4$~~~~~&~~~~~$20.6$~~~~~&~~~~~$1.803$~~~~~&~~~~~$0.984$~~~~~&~~~~~$27.4\pm5.1\pm1.3$~~~~~\\
~~~~~$3.0800$~~~~~&~~~~~$157\pm1$~~~~~&~~~~~$298.5\pm17.6$~~~~~&~~~~~$19.4$~~~~~&~~~~~$1.901$~~~~~&~~~~~$0.915$~~~~~&~~~~~$29.1\pm1.7\pm1.2$~~~~~\\
  \hline\hline
  \end{tabular}}
  \label{table_XS_result}
  \end{center}
\end{table*}

\subsection{\boldmath Systematic uncertainties}
Several sources of systematic uncertainties are considered in the determination of $\sigma_{\phi\eta}^{{\rm{Born}}}$. 
The uncertainties associated with the knowledge of the tracking efficiency of the two charged tracks as well as from the PID efficiency are studied with a $e^{+}e^{-}\rightarrow K^{+}K^{-}\pi^{+}\pi^{-}$ control sample.
The difference of the efficiency measured in data and MC is assigned as the uncertainty, and it is found to be 1.0\% per track for both tracking and PID efficiency~\cite{Dong}.
The uncertainty due to photon reconstruction efficiency is 1.0\% per photon~\cite{gamsys}.
The uncertainty of the luminosity measurement is smaller than 1.0\%~\cite{cite_lumi}.
The uncertainty associated with the kinematic fit is estimated by not using the correction of the helix parameters of the charged tracks described in detail in Ref.~\cite{kinsys} and taking the difference to the nominal result as the uncertainty.
In order to estimate the contribution from the $\eta$ selection, the mass window is varied from $|M(\gamma\gamma)-m_{\eta}| <3\sigma$ to 2.5$\sigma$ and 3.5$\sigma$, and the larger difference to the nomimal result is taken as the uncertainty.
The systematic uncertainty of the signal yield is estimated by varying the fit range from (0.98, 1.08)~GeV/$c^{2}$ to (0.99, 1.09)~GeV/$c^{2}$, where the difference to the nominal result is the uncertainty.
The uncertainty related to the signal shape is estimated with an alternative fit using the $\phi$ MC shape convolved with a Gaussian function.
The uncertainty due to background shape is estimated with an alternative fit using an Argus function~\cite{ARGUS} instead of a polynomial.
The uncertainty due to $\phi$ peaking background is estimated by investigating the $\eta$ sideband which is defined as $40$~MeV$/c^2< |M(\gamma\gamma)-m_{\eta}| <150$~MeV$/c^2$. We take the difference of the normalized number of the events from a signal fit to the sideband region and the number of events estimated from MC as systematic uncertainties. 
%
The uncertainty in $\epsilon(1+\delta)$ arises from the accuracy of the radiator function, which is about 0.5\%~\cite{effisr}, and has an additional contribution from the parametrization of the $\sigma_{\phi\eta}^{{\rm{Born}}}(s)$ line shape, which is taken as the standard deviation of the fit to the sampled parameters.
The two contributions are summed in quadrature.
The uncertainties of the branching fractions of intermediate states are 1.1\%~\cite{PDG}.
Assuming that these contributions to the systematic uncertainties are uncorrelated, the total systematic uncertainties are obtained by adding the individual uncertainties in quadrature.
The resulting values for all $\sqrt{s}$ are shown in Table~\ref{tab::sys}.
The fluctuations of some relative uncertainties among the different energies origin from the influence of statistics, and have negligible effect on the total absolute uncertainties and the final results.
The total systematic uncertainties on $\sigma_{\phi\eta}^{{\rm{Born}}}$ range from 3.9\% to 8.3\%.

 \begin{table*}[htbp]
  \begin{center}
  \caption{
Relative systematic uncertainties (in \%) of the measurement of $\sigma_{\phi\eta}^{{\rm{Born}}}$ from the tracking efficiency (Trk), PID efficiency (PID), photon detection ($\gamma$ detect), luminosity (Lum), the kinematic fit (KinFit), $\eta$ selection ($\eta$ range), fit range (Fit range), signal shape (Signal), background shape (Background), $\phi$ peaking background (Side), radiative correction ($\epsilon(1+\delta)$) and branching fractions ($\mathcal{B}$). The total uncertainty is obtained by summing the individual contributions in quadrature.
  }
  \resizebox{!}{4.39cm}{\begin{tabular}{c c c c c c c c c c c c c c }
      \hline \hline
        $\sqrt{s}$~[GeV]& Trk & PID & $\gamma$ detect & Lum & KinFit & $\eta$ range & Fit range & Signal & Background & Side & $\epsilon(1+\delta)$ &$\mathcal{B}$ & Total\\ \hline

~~~~~$2.0000$~~~&~~~$2.0$~~~&~~~$2.0$~~~&~~~$2.0$~~~&~~~$0.7$~~~&~~~$0.0$~~~&~~~$1.0$~~~&~~~$0.0$~~~&~~~$0.5$~~~&~~~$1.1$~~~&~~~$0.5$~~~&~~~$0.5$~~~&~~~$1.1$~~~&~~~$4.1$~~~~~\\
~~~~~$2.0500$~~~&~~~$2.0$~~~&~~~$2.0$~~~&~~~$2.0$~~~&~~~$0.8$~~~&~~~$1.4$~~~&~~~$1.3$~~~&~~~$1.5$~~~&~~~$0.0$~~~&~~~$0.1$~~~&~~~$0.0$~~~&~~~$0.5$~~~&~~~$1.1$~~~&~~~$4.5$~~~~~\\
~~~~~$2.1000$~~~&~~~$2.0$~~~&~~~$2.0$~~~&~~~$2.0$~~~&~~~$0.7$~~~&~~~$1.2$~~~&~~~$0.6$~~~&~~~$0.1$~~~&~~~$0.5$~~~&~~~$1.0$~~~&~~~$1.1$~~~&~~~$0.5$~~~&~~~$1.1$~~~&~~~$4.3$~~~~~\\
~~~~~$2.1250$~~~&~~~$2.0$~~~&~~~$2.0$~~~&~~~$2.0$~~~&~~~$0.9$~~~&~~~$0.8$~~~&~~~$0.3$~~~&~~~$0.0$~~~&~~~$0.2$~~~&~~~$0.1$~~~&~~~$0.4$~~~&~~~$0.5$~~~&~~~$1.1$~~~&~~~$3.9$~~~~~\\
~~~~~$2.1500$~~~&~~~$2.0$~~~&~~~$2.0$~~~&~~~$2.0$~~~&~~~$0.9$~~~&~~~$0.2$~~~&~~~$1.5$~~~&~~~$0.2$~~~&~~~$0.9$~~~&~~~$0.6$~~~&~~~$0.0$~~~&~~~$0.5$~~~&~~~$1.1$~~~&~~~$4.2$~~~~~\\
~~~~~$2.1750$~~~&~~~$2.0$~~~&~~~$2.0$~~~&~~~$2.0$~~~&~~~$0.9$~~~&~~~$0.4$~~~&~~~$2.8$~~~&~~~$0.1$~~~&~~~$0.8$~~~&~~~$1.0$~~~&~~~$1.2$~~~&~~~$1.6$~~~&~~~$1.1$~~~&~~~$5.3$~~~~~\\
~~~~~$2.2000$~~~&~~~$2.0$~~~&~~~$2.0$~~~&~~~$2.0$~~~&~~~$0.7$~~~&~~~$0.5$~~~&~~~$0.8$~~~&~~~$0.3$~~~&~~~$0.6$~~~&~~~$1.3$~~~&~~~$0.0$~~~&~~~$0.5$~~~&~~~$1.1$~~~&~~~$4.1$~~~~~\\
~~~~~$2.2324$~~~&~~~$2.0$~~~&~~~$2.0$~~~&~~~$2.0$~~~&~~~$0.6$~~~&~~~$0.6$~~~&~~~$1.1$~~~&~~~$2.3$~~~&~~~$1.3$~~~&~~~$1.1$~~~&~~~$1.3$~~~&~~~$0.5$~~~&~~~$1.1$~~~&~~~$5.0$~~~~~\\
~~~~~$2.3094$~~~&~~~$2.0$~~~&~~~$2.0$~~~&~~~$2.0$~~~&~~~$0.7$~~~&~~~$0.1$~~~&~~~$1.2$~~~&~~~$0.0$~~~&~~~$0.0$~~~&~~~$0.1$~~~&~~~$1.5$~~~&~~~$0.5$~~~&~~~$1.1$~~~&~~~$4.2$~~~~~\\
~~~~~$2.3864$~~~&~~~$2.0$~~~&~~~$2.0$~~~&~~~$2.0$~~~&~~~$0.8$~~~&~~~$0.7$~~~&~~~$0.7$~~~&~~~$1.3$~~~&~~~$1.0$~~~&~~~$0.2$~~~&~~~$0.5$~~~&~~~$0.5$~~~&~~~$1.1$~~~&~~~$4.3$~~~~~\\
~~~~~$2.3960$~~~&~~~$2.0$~~~&~~~$2.0$~~~&~~~$2.0$~~~&~~~$0.7$~~~&~~~$0.8$~~~&~~~$1.1$~~~&~~~$0.4$~~~&~~~$0.7$~~~&~~~$1.3$~~~&~~~$0.8$~~~&~~~$0.5$~~~&~~~$1.1$~~~&~~~$4.4$~~~~~\\
~~~~~$2.4000$~~~&~~~$2.0$~~~&~~~$2.0$~~~&~~~$2.0$~~~&~~~$0.7$~~~&~~~$0.2$~~~&~~~$1.8$~~~&~~~$2.7$~~~&~~~$0.7$~~~&~~~$0.9$~~~&~~~$0.0$~~~&~~~$0.5$~~~&~~~$1.1$~~~&~~~$5.1$~~~~~\\
~~~~~$2.6444$~~~&~~~$2.0$~~~&~~~$2.0$~~~&~~~$2.0$~~~&~~~$0.6$~~~&~~~$0.1$~~~&~~~$0.6$~~~&~~~$0.5$~~~&~~~$0.4$~~~&~~~$0.8$~~~&~~~$0.3$~~~&~~~$0.5$~~~&~~~$1.1$~~~&~~~$3.9$~~~~~\\
~~~~~$2.6464$~~~&~~~$2.0$~~~&~~~$2.0$~~~&~~~$2.0$~~~&~~~$0.8$~~~&~~~$0.0$~~~&~~~$1.3$~~~&~~~$0.1$~~~&~~~$0.7$~~~&~~~$0.5$~~~&~~~$1.1$~~~&~~~$0.5$~~~&~~~$1.1$~~~&~~~$4.2$~~~~~\\
~~~~~$2.9000$~~~&~~~$2.0$~~~&~~~$2.0$~~~&~~~$2.0$~~~&~~~$0.9$~~~&~~~$0.4$~~~&~~~$0.9$~~~&~~~$0.2$~~~&~~~$0.0$~~~&~~~$0.9$~~~&~~~$2.7$~~~&~~~$0.5$~~~&~~~$1.1$~~~&~~~$4.9$~~~~~\\
~~~~~$2.9500$~~~&~~~$2.0$~~~&~~~$2.0$~~~&~~~$2.0$~~~&~~~$0.9$~~~&~~~$0.4$~~~&~~~$1.8$~~~&~~~$0.3$~~~&~~~$0.8$~~~&~~~$0.4$~~~&~~~$0.0$~~~&~~~$0.5$~~~&~~~$1.1$~~~&~~~$4.3$~~~~~\\
~~~~~$2.9810$~~~&~~~$2.0$~~~&~~~$2.0$~~~&~~~$2.0$~~~&~~~$0.6$~~~&~~~$0.7$~~~&~~~$1.8$~~~&~~~$3.5$~~~&~~~$0.9$~~~&~~~$0.6$~~~&~~~$0.0$~~~&~~~$0.5$~~~&~~~$1.1$~~~&~~~$5.6$~~~~~\\
~~~~~$3.0000$~~~&~~~$2.0$~~~&~~~$2.0$~~~&~~~$2.0$~~~&~~~$0.7$~~~&~~~$1.1$~~~&~~~$1.8$~~~&~~~$1.8$~~~&~~~$0.7$~~~&~~~$0.6$~~~&~~~$5.2$~~~&~~~$0.5$~~~&~~~$1.1$~~~&~~~$7.0$~~~~~\\
~~~~~$3.0200$~~~&~~~$2.0$~~~&~~~$2.0$~~~&~~~$2.0$~~~&~~~$0.7$~~~&~~~$4.0$~~~&~~~$6.3$~~~&~~~$0.5$~~~&~~~$0.0$~~~&~~~$0.0$~~~&~~~$0.0$~~~&~~~$0.5$~~~&~~~$1.1$~~~&~~~$8.3$~~~~~\\
~~~~~$3.0500$~~~&~~~$2.0$~~~&~~~$2.0$~~~&~~~$2.0$~~~&~~~$0.7$~~~&~~~$0.0$~~~&~~~$1.7$~~~&~~~$0.9$~~~&~~~$0.0$~~~&~~~$0.1$~~~&~~~$0.0$~~~&~~~$0.5$~~~&~~~$1.1$~~~&~~~$4.2$~~~~~\\
~~~~~$3.0600$~~~&~~~$2.0$~~~&~~~$2.0$~~~&~~~$2.0$~~~&~~~$0.9$~~~&~~~$0.7$~~~&~~~$2.3$~~~&~~~$0.0$~~~&~~~$1.3$~~~&~~~$1.1$~~~&~~~$0.0$~~~&~~~$0.5$~~~&~~~$1.1$~~~&~~~$4.8$~~~~~\\
~~~~~$3.0800$~~~&~~~$2.0$~~~&~~~$2.0$~~~&~~~$2.0$~~~&~~~$0.6$~~~&~~~$0.6$~~~&~~~$0.5$~~~&~~~$0.6$~~~&~~~$0.2$~~~&~~~$0.8$~~~&~~~$1.4$~~~&~~~$0.5$~~~&~~~$1.1$~~~&~~~$4.2$~~~~~\\     \hline\hline
  \end{tabular}}
  \label{tab::sys}
  \end{center}
\end{table*}

\section{Line shape of $e^{+}e^{-}\rightarrow\phi\eta$}
\label{lineshape}
\subsection{Fit of the line shape}
To study a possible resonant behavior in $e^{+}e^{-}\rightarrow\phi\eta$, a least $\chi^{2}$ fit taking into account the correlation between systematic uncertainties for different ($\sqrt{s}$) is performed to the measured values of the Born cross section $\sigma_{\phi\eta}^{{\rm{Born}}}$.
Previous results from the {\textsl{BABAR}} collaborations~\cite{Y2175BABAR3} are also included to be able to describe the low-energy behavior of the cross section.
Following Ref.~\cite{Y2175BABAR3} and assuming that the reaction proceeds mostly through the decay of the two resonances $\phi(1680)$ and $\phi(2170)$, 
the line shape is fitted using a coherent sum of two phase-space factor modified Breit-Wigner functions and a non-resonant term:
\begin{eqnarray}
         \label{eq:FitSec}
         \begin{split}
         \sigma_{\phi\eta}^{{\rm{Born}}}(s) =& 12\mathcal{\pi}\mathcal{P}_{\phi\eta}(s)\left|A_{\phi\eta}^{{\rm{n.r.}}}(s) \right.\\
         &\left.+A_{\phi\eta}^{\phi(1680)}(s)+A_{\phi\eta}^{\phi(2170)}(s)\right|^{2},
         \end{split}
         \end{eqnarray}
where $\mathcal{P}_{\phi\eta}(s)$ is the phase space factor of the $\phi\eta$ system, $A_{\phi\eta}^{{\rm{n.r.}}}(s)$ describes the non-resonant contribution, 
$A_{\phi\eta}^{\phi(1680)}(s)$ and $A_{\phi\eta}^{\phi(2170)}(s)$ are the two vector resonances.
For the phase space for the $\phi\eta$ system, we use
\begin{eqnarray}
         \label{eq:Phase}
         \mathcal{P}_{\phi\eta}(s)=\left[\frac{(s+m_{\phi}^{2}-m_{\eta}^{2})^{2}-4m_{\phi}^{2}s}{s}\right]^{\frac{3}{2}}.
         \end{eqnarray}
A reasonable description of the non-resonant contribution is given by the power-law dependence $A_{\phi\eta}^{{\rm{n.r.}}}(s)=a_{0}/s^{a_{1}}$.
We describe the $\phi(1680)$ resonance using Breit-Wigner amplitude
\begin{eqnarray}
         \label{eq:FitGamY1685}
         \begin{split}
         A_{\phi\eta}^{\phi(1680)}(s)=&\sqrt{\mathcal{B}_{\phi\eta}^{\phi(1680)}\Gamma_{e^{+}e^{-}}^{\phi(1680)}} \times
         \\
         &\frac{\sqrt{\frac{\Gamma_{\phi(1680)}}{\mathcal{P}_{\phi\eta}(m_{\phi(1680)}^{2})}}}{m_{\phi(1680)}^{2}-s-\mathit{i}\sqrt{s}\Gamma_{\phi(1680)}(s)},
         \end{split}
         \end{eqnarray}
with partial width $\mathcal{B}_{\phi\eta}^{\phi(1680)}\Gamma_{e^{+}e^{-}}^{\phi(1680)}$, mass $m_{\phi(1680)}$, width $\Gamma_{\phi(1680)}$.
The width $\Gamma_{\phi(1680)}$ on the denominator use an energy dependent width~\cite{Y2175BABAR3}
\begin{eqnarray}
         \label{eq:FitGamtot}
         \begin{split}
         \Gamma_{\phi(1680)}(s)=&\Gamma_{\phi(1680)}\left[\frac{\mathcal{P}_{\phi\eta}(s)}{\mathcal{P}_{\phi\eta}(m_{\phi(1680)}^{2})}\mathcal{B}_{\phi\eta}^{\phi(1680)} \right. \\
         &\left.+(1-\mathcal{B}_{\phi\eta}^{\phi(1680)})\right].
         \end{split}
         \end{eqnarray}
In the fit, the previous results from the {\textsl{BABAR}} collaboration~\cite{Y2175BABAR3} are included to be able to also describe the low-energy behavior ($<$2~GeV) of the cross section where we have no data. 
The parameters of the $\phi(1680)$, such as mass ($m_{\phi(1680)}=1709$~MeV/$c^{2}$), width ($\Gamma_{\phi(1680)}=369$~MeV) and branching fraction ($\Gamma_{e^{+}e^{-}}^{\phi(1680)}\mathcal{B}_{\phi\eta}^{\phi(1680)}=138$~eV), are fixed to the results from Ref.~\cite{Y2175BABAR3}.
The $\phi(2170)$ is described using Breit-Wigner amplitude
\begin{eqnarray}
         \label{eq:FitGamY2175}
         \begin{split}
         A_{\phi\eta}^{\phi(2170)}(s)=&\sqrt{\mathcal{B}_{\phi\eta}^{\phi(2170)}\Gamma_{e^{+}e^{-}}^{\phi(2170)}} \times
         \\
         &\frac{\sqrt{\frac{\Gamma_{\phi(2170)}}{\mathcal{P}_{\phi\eta}(m_{\phi(2170)}^{2})}}\mathit{e}^{\mathit{i}\Psi_{\phi(2170)}}}{m_{\phi(2170)}^{2}-s-\mathit{i}\sqrt{s}\Gamma_{\phi(2170)}}\cdot\frac{B(p)}{B(p')},
         \end{split}
         \end{eqnarray}
with partial width $\mathcal{B}_{\phi\eta}^{\phi(2170)}\Gamma_{e^{+}e^{-}}^{\phi(2170)}$, mass $m_{\phi(2170)}$, width $\Gamma_{\phi(2170)}$, and relative phase angle to the non-resonant component $\Psi_{\phi(2170)}$. $B(p)$ is the $P$-wave Blatt-Weisskopf form factor, $p$ is the breakup momentum corresponding to the ($\sqrt{s}$) and $p’$ is the breakup momentum at the $\phi(2170)$ mass.

\begin{figure*}[htbp]
\begin{center}
\centering
\vskip-20pt
\mbox{
\hskip-6pt
  \begin{overpic}[width=8.5cm,height=6cm,angle=0]{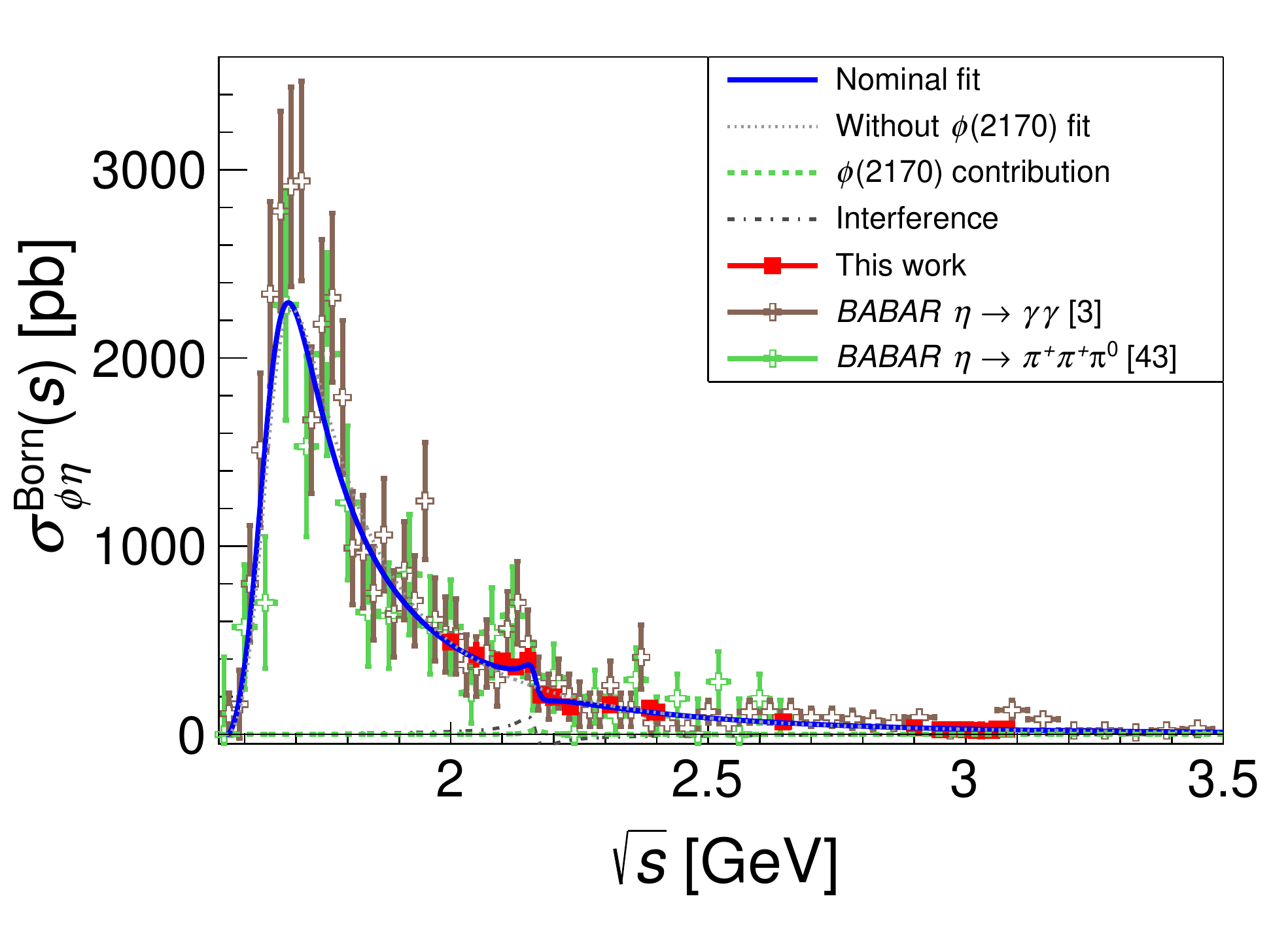}
  \put(0,70){$(a)$}
  \end{overpic}
 \begin{overpic}[width=8.5cm,height=6cm,angle=0]{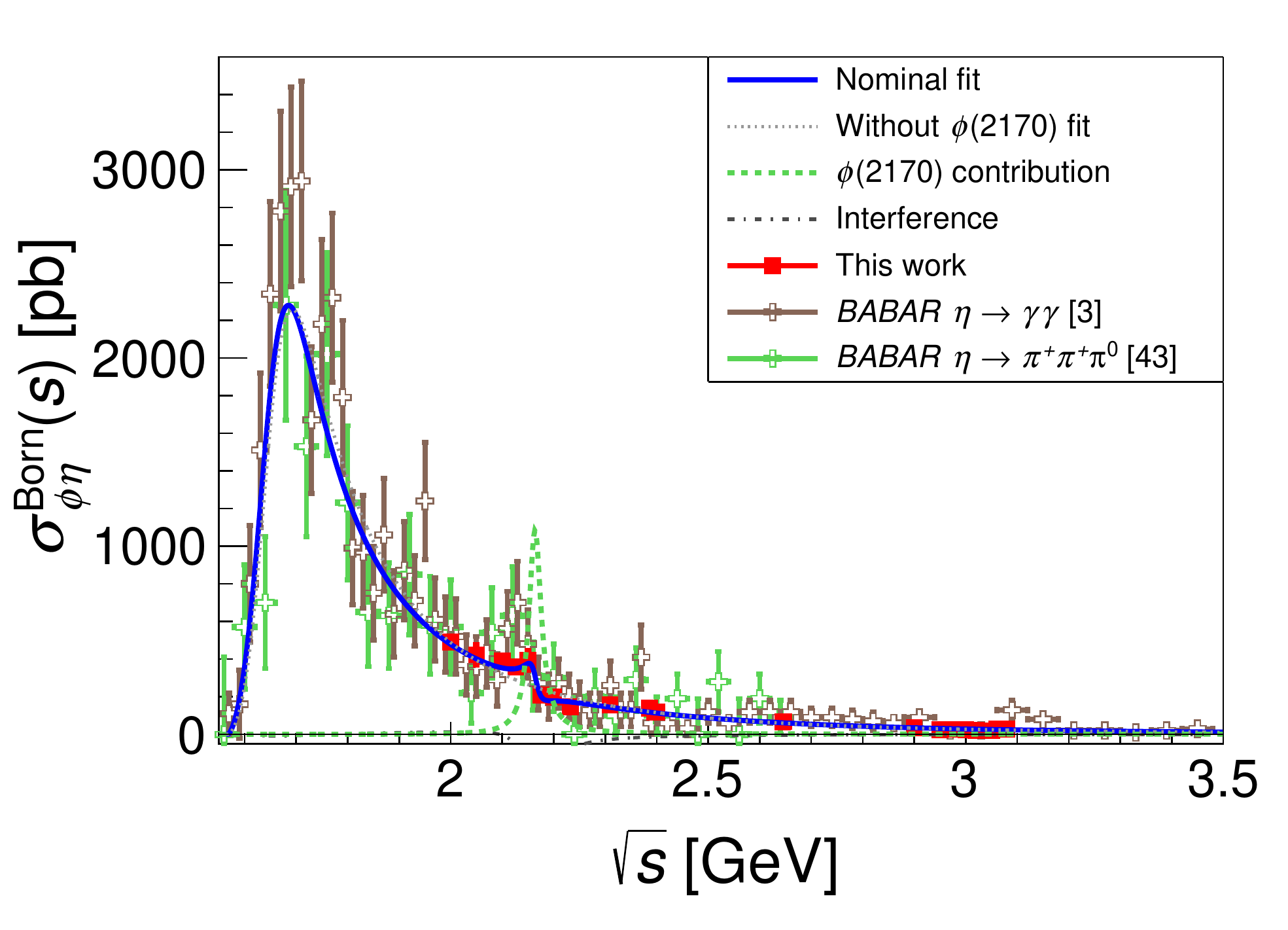}
  \put(0,70){$(b)$}
  \end{overpic}
 }
 \mbox{
\hskip-6pt
  \begin{overpic}[width=8.5cm,height=6cm,angle=0]{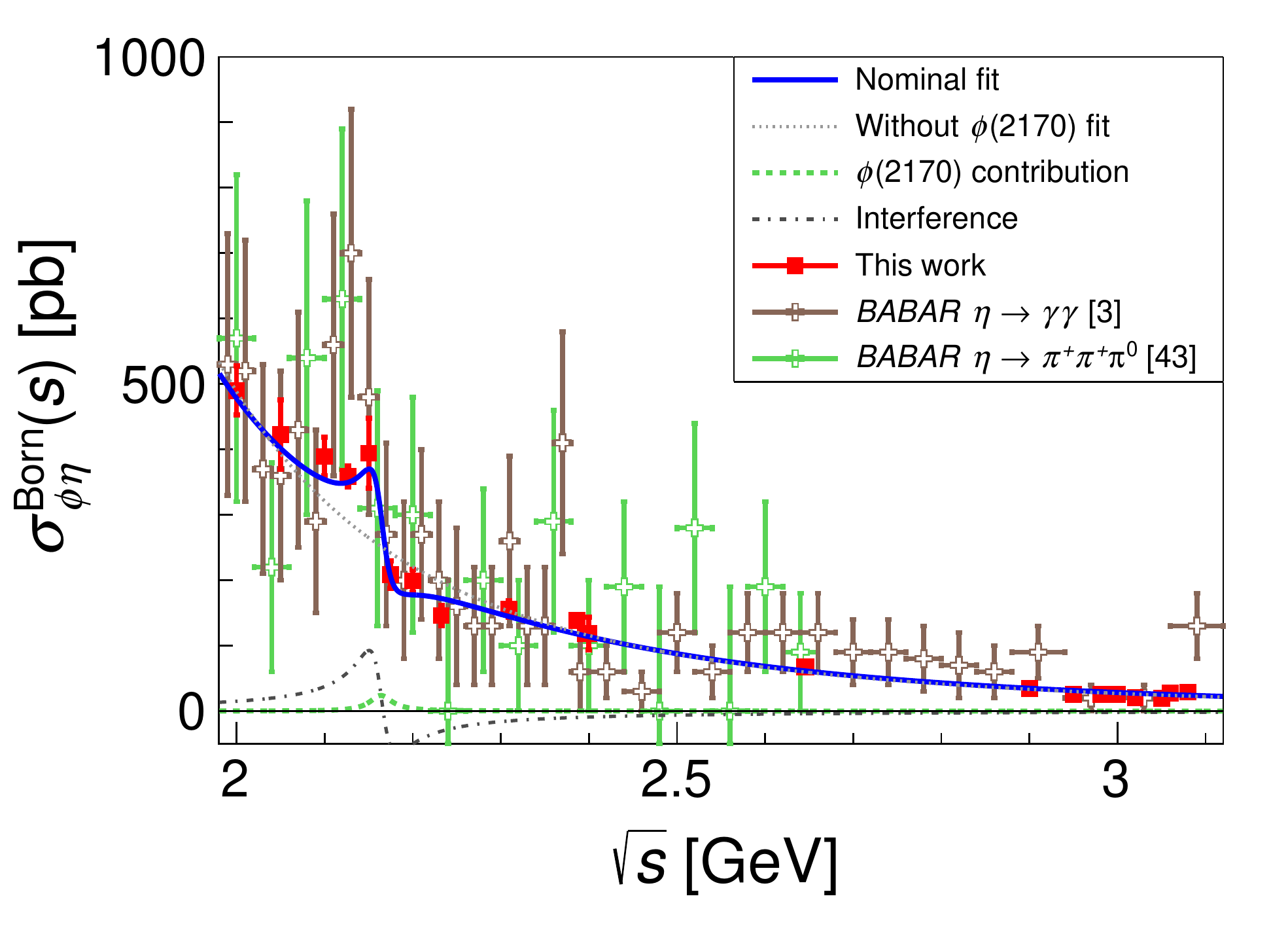}
  \put(0,70){$(c)$}
  \end{overpic}
 \begin{overpic}[width=8.5cm,height=6cm,angle=0]{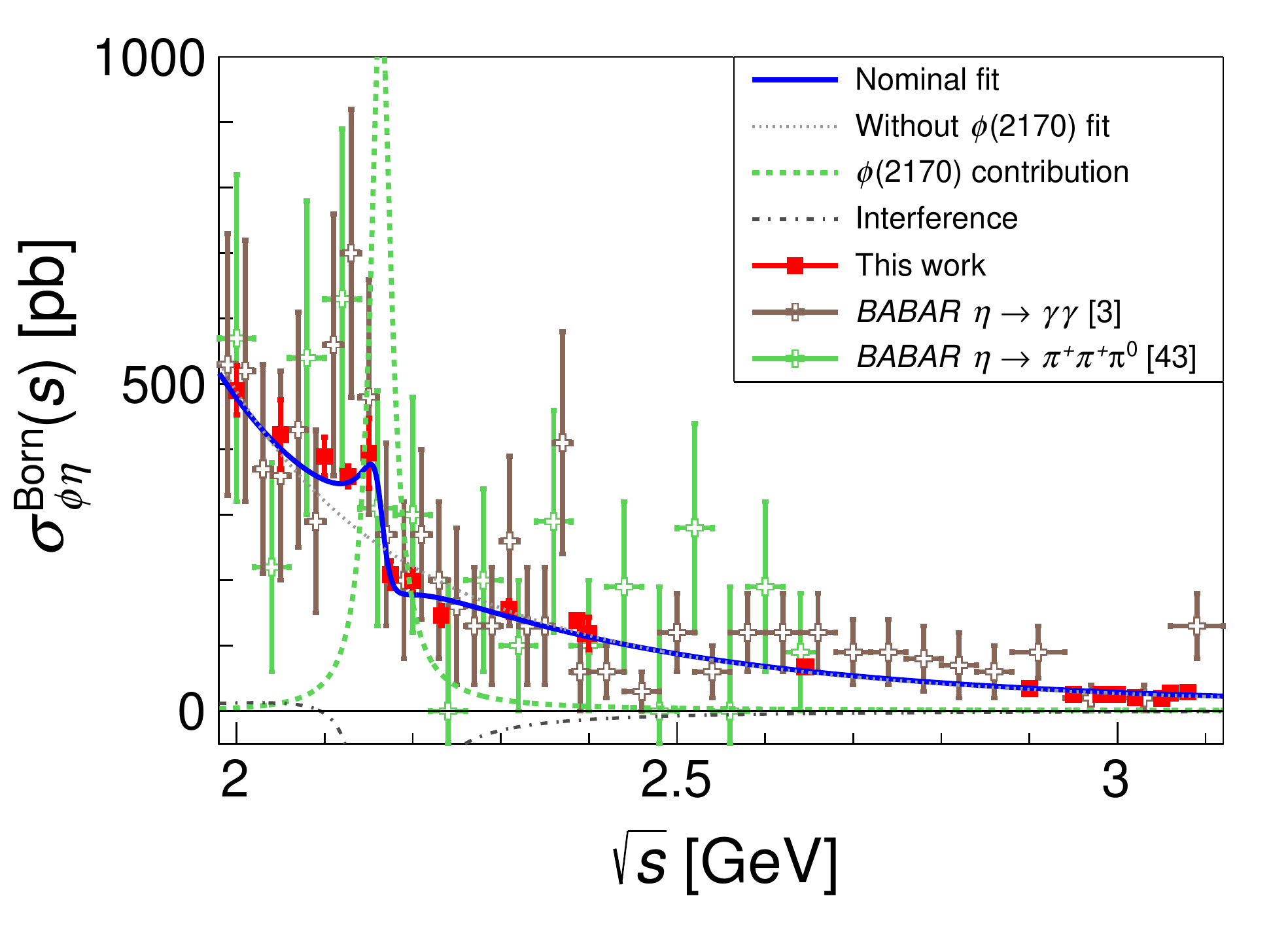}
  \put(0,70){$(d)$}
  \end{overpic}
 }
 \mbox{
\hskip-6pt
  \begin{overpic}[width=8.5cm,height=6cm,angle=0]{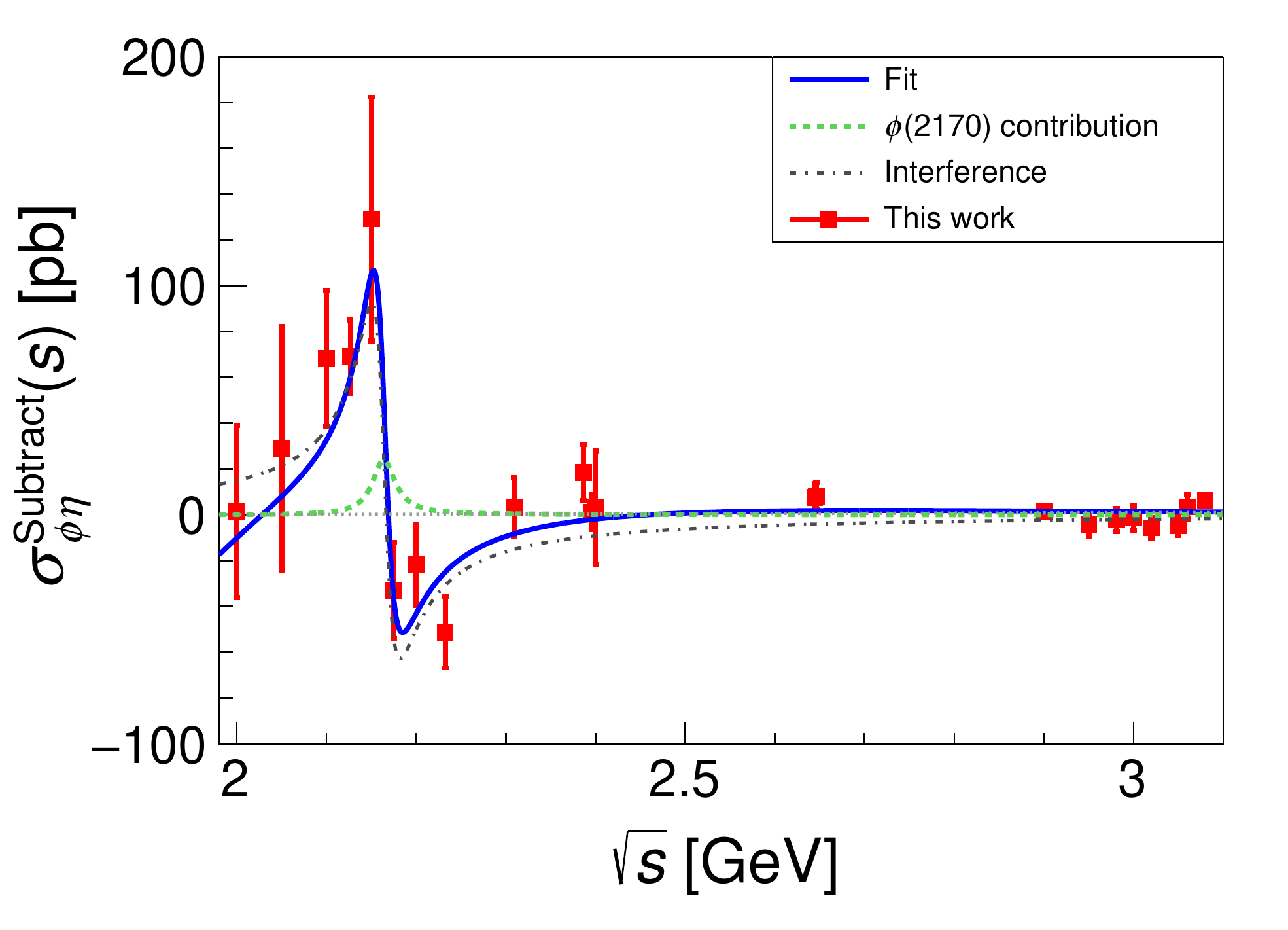}
  \put(0,70){$(e)$}
  \end{overpic}
 \begin{overpic}[width=8.5cm,height=6cm,angle=0]{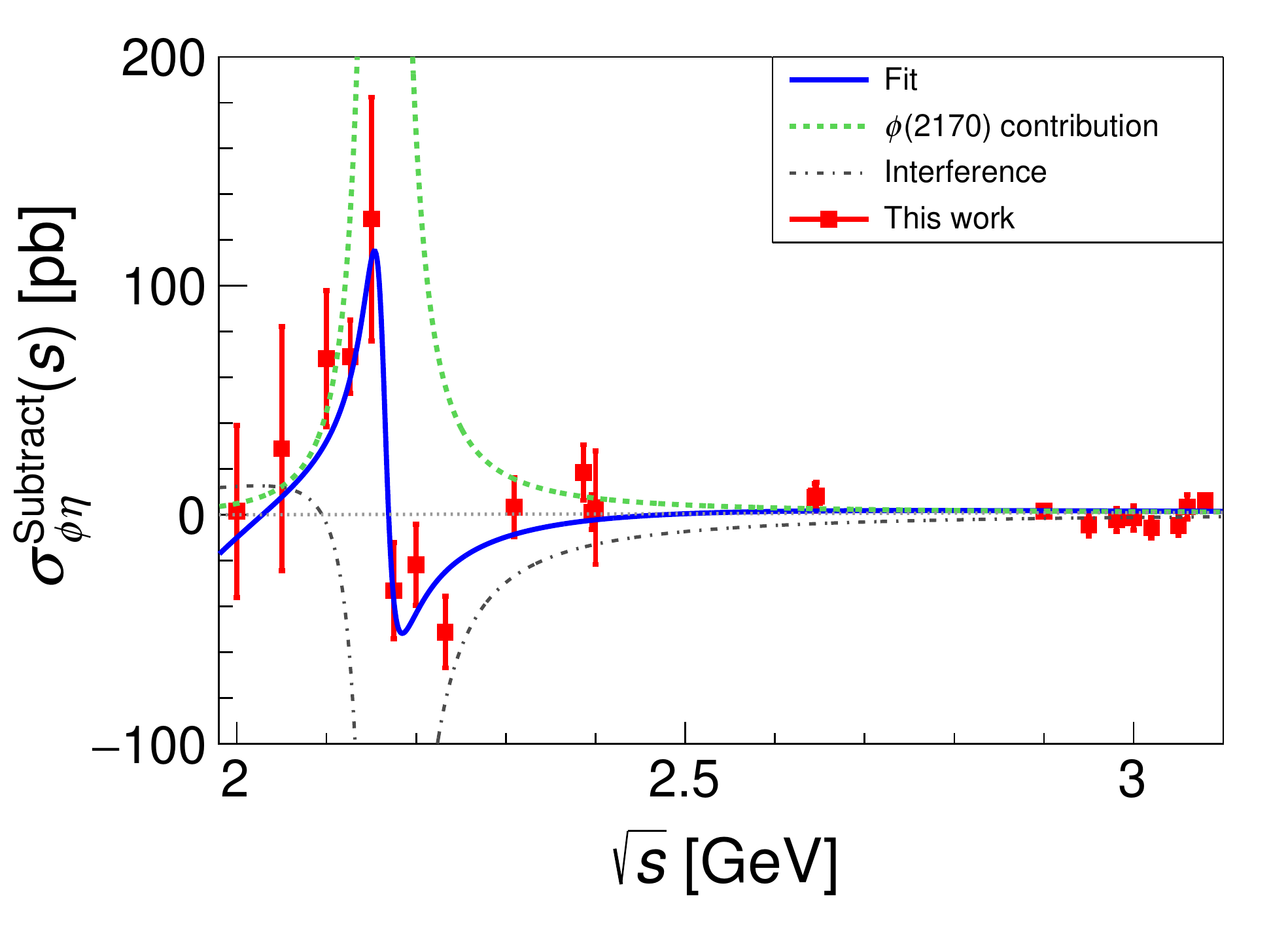}
  \put(0,70){$(f)$}
  \end{overpic}
 }
 \vskip+50.0pt
 \caption{Results from this work (red solid squares) including statistical and systematic uncertainties
   for the $e^{+}e^{-}\rightarrow\phi\eta$ cross section and a fit to the combined data of {\textsl{BABAR} and this work} (blue solid line) (a) (c) Solution~I. (b) (d) Solution~II.;
   cross section and the nominal fit after subtraction of the fit curve without inclusion of the $\phi(2170)$ (e) Solution~I. (f) Solution~II.
   Also shown are previously published measurements from {\textsl{BABAR}} ($\eta\rightarrow\gamma\gamma$)~\cite{Y2175BABAR3} and {\textsl{BABAR}} ($\eta\rightarrow\pi^{+}\pi^{-}\pi^{0}$)~\cite{Y2175phietap}.}
\label{fig_Xsec}
\end{center}
\end{figure*}

The fit has two solutions with an identical mass and width of the resonance.
The fit quality is estimated by the $\chi^{2}$, where the best fit
gives a $\chi^{2}/{\rm{n.d.f}}=0.9$, with $\rm{n.d.f}=97$ being the
number of degrees of freedom.  The mass and width of
$\phi(2170)$ are determined to be
$m_{\phi(2170)}=(2163.5\pm6.2)$~MeV/$c^{2}$ and $\Gamma_{\phi(2170)}=(31.1_{-11.6}^{+21.1})$~MeV
from our fit, illustrated in Fig.~\ref{fig_Xsec}. 
The significance of the $\phi(2170)$ resonance is determined to
be 6.9$\sigma$. This is obtained by
comparing the change of $\Delta\chi^{2} = 59.05$ with (blue solid line in
Fig.~\ref{fig_Xsec}~(a)$\sim$(d)) and without (grey dotted line in
Fig.~\ref{fig_Xsec}~(a)$\sim$(d)) the resonance in the fit and taking the
change of number of degree of freedom $\Delta{\rm{n.d.f}}=4$ into
account.  The cross section and the fit results are summarized in Table~~\ref{table_Fit_result} and shown in
Fig.~\ref{fig_Xsec}. Fig.~\ref{fig_Xsec}~(a) and (c) are solution~I; Fig.~\ref{fig_Xsec}~(b) and (d) are solution~II.
Fig.~\ref{fig_Xsec}~(b) shows the same data
subtracting the fit result that is obtained without inclusion of the
$\phi(2170)$. It is obvious that an additional resonant
structure around 2.175~GeV is needed.

\begin{table}[htbp]
  \begin{center}
  \caption{
Parameters for resonances $\phi(2170)$ obtained by fitting 
  }
  \resizebox{!}{1.9cm}{\begin{tabular}{c c c}
      \hline \hline
      Parameter & Solution~I & Solution~II\\\hline 
~~~$\chi^{2}/{\rm{n.d.f}}$~~~&\multicolumn{2}{c}{~~~$86.8/97$~~~}\\
~~~$a_0$~~~&~~~$-0.11_{-0.22}^{+0.09}$~~~&~~~$-0.24\pm0.58$~~~\\
~~~$a_1$~~~&~~~$1.91_{-0.76}^{+0.61}$~~~&~~~$2.54_{-1.55}^{+2.86}$~~~\\
~~~$\mathcal{B}_{\phi\eta}^{\phi(2170)}\Gamma_{e^{+}e^{-}}^{\phi(2170)}$~~~&~~~$0.24_{-0.07}^{+0.12}$~eV~~~&~~~$10.11_{-3.13}^{+3.87}$~eV~~~\\
~~~$m_{\phi(2170)}$~~~&\multicolumn{2}{c}{~~~$2163.5\pm6.2$~MeV/$c^{2}$~~~}\\
~~~$\Gamma_{\phi(2170)}$~~~&\multicolumn{2}{c}{~~~$31.1_{-11.6}^{+21.1}$~MeV~~~}\\
~~~$\Phi_{\phi(2170)}$~~~&~~~$1.82_{-0.31}^{+0.35}$~~~&~~~$-2.92_{-0.06}^{+0.05}$~~~\\\hline\hline
  \end{tabular}}
  \label{table_Fit_result}
  \end{center}
\end{table}

\subsection{\boldmath Systematic uncertainties}
For the systematic uncertainties of the resonance parameters, we
examine effects from the choice of the model for the non-resonant
contribution and of the fit range.  For the model dependence of the
non-resonant contribution, a $A_{\phi\eta}^{\rm{n.r.}}(s)=a_{0}/s$ is
used instead, resulting in differences of 3.0~MeV/$c^{2}$ and 0.1~MeV for mass and width, respectively.  For the
dependence on the fit range, we set aside the energy
point 2.00 and 3.08 GeV, resulting in differences of 0.3~MeV/$c^{2}$ and 1.1~MeV for mass and width, respectively.
The total systematic uncertainties are thus 3.0~MeV/$c^{2}$, 1.1~MeV
for mass and width, respectively.

\begin{figure}[htbp]
\begin{center}
\begin{overpic}[width=8.5cm,height=8.0cm,angle=0]{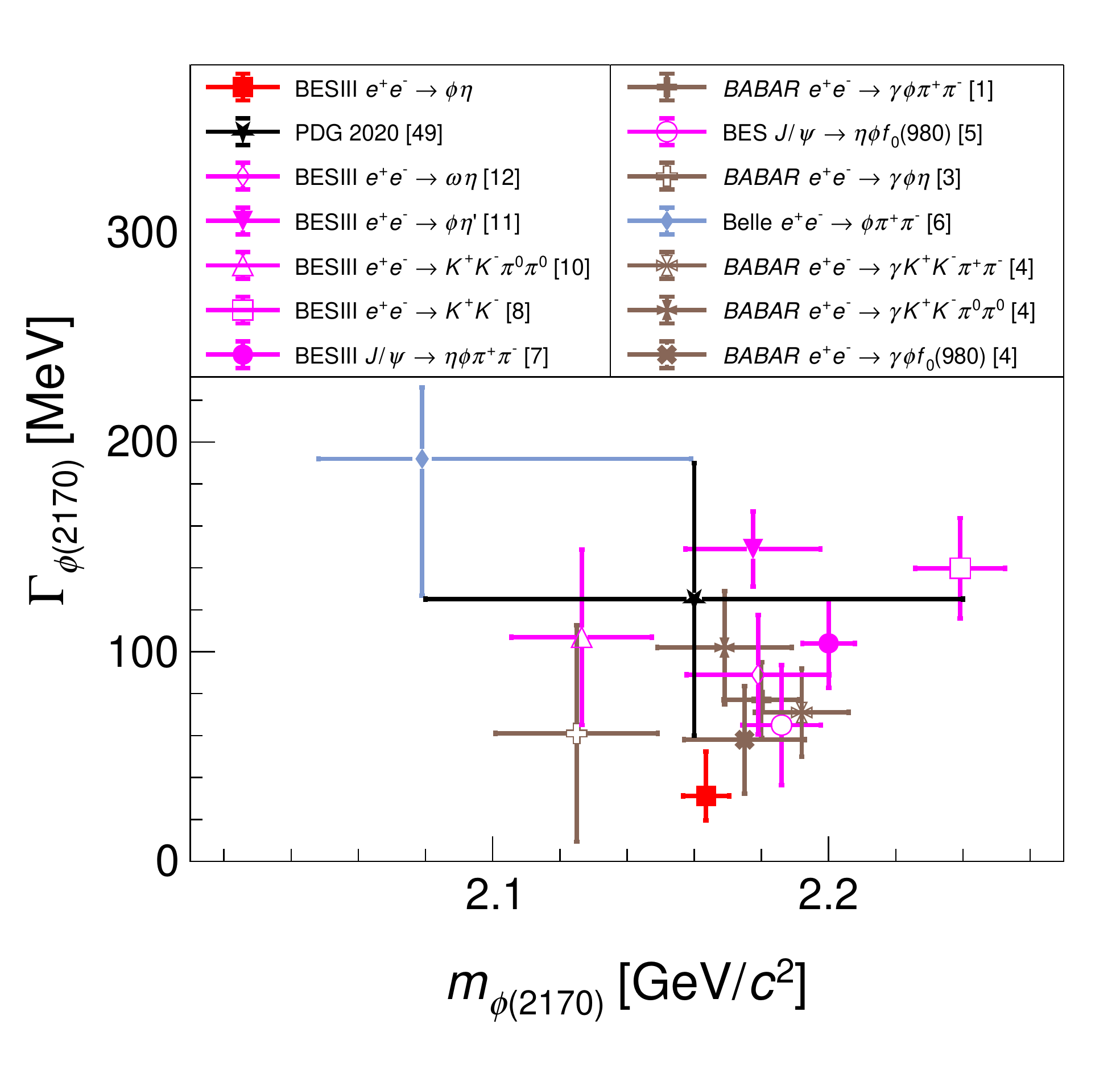}
\end{overpic}
\caption{The parameters of the $\phi(2170)$ state obtained from different processes compared to those obtained in this work in the $e^{+}e^{-} \rightarrow \phi\eta$ process.}
\label{fig::MGcom}
\end{center}
\end{figure}

\section{\boldmath Summary and Discussion}
This paper presents the most accurate measurement of the Born cross
section of $e^{+}e^{-} \rightarrow \phi\eta$, at 22 c.m. energies in
the interval 2.00 to 3.08 GeV.  A resonant structure is observed in
the $\sigma_{\phi\eta}^{{\rm{Born}}}$ line shape.  We determine the parameters of this resonance to be
$m_{\phi(2170)}=(2163.5\pm6.2\pm3.0)$~MeV/$c^{2}$ and
$\Gamma_{\phi(2170)}=(31.1_{-11.6}^{+21.1}\pm1.1)$~MeV. Here, the first
uncertainties are statistical and the second ones are systematic.  The significance is larger than 6.9$\sigma$. 
With the input of the partial width of $\mathcal{B}_{\phi\eta'}^{R}\Gamma_{ee}^{R}=(7.1\pm0.7)$~eV~\cite{Yankun2}, the ratio of partial widths between the decay modes $\phi\eta$ and $\phi\eta'$ would be ($0.03_{-0.01}^{+0.02}$) or ($1.42_{-0.46}^{+0.56}$).
Compared to a previous measurement by the {\textsl{BABAR}}
experiment~\cite{Y2175BABAR3}, the mass value reported here is
significantly larger.  While similar resonances have been observed in
many different channels, the observed decay widths vary significantly.

The fitted result is compared with the parameters of the $\phi(2170)$ state measured by previous experiments via various processes as shown in Fig.~\ref{fig::MGcom}. 
The results obtained in this paper are consistent with the world average parameters of the $\phi(2170)$. However, differences to other individual measurements in different channels can be sizable.
Among the existing measurements, the result of this measurement yields the smallest width of the $\phi(2170)$ resonance observed so far.


In principle, the resonance observed in this work could also be an excited $\omega$. However, according to the OZI rule~\cite{OZI_rule1,OZI_rule2,OZI_rule3}, intermediate $\omega$-like resonances should be highly suppressed in the $\phi\eta$ channel.
The widths of vector states $\omega(2205)$, $\omega(2290)$ and $\omega(2330)$ listed in PDG~\cite{PDG} are much larger than the results in $\phi\eta$ mode.
The mass and width measured in this work do not agree with those of the $\rho$-like resonance found in $e^{+}e^{-}\to\omega\pi^{0}$ by BESIII  ($m_{\phi(2170)}=(2034\pm13\pm9)$~MeV/$c^{2}$, $\Gamma_{\phi(2170)}=(234\pm30\pm25)$~MeV)~\cite{Dong2}.
$\rho$-like contributions should be suppressed due to the isoscalar nature of the $\phi\eta$ channel.
Our results agree with those of $\rho$-like resonance found in a recent work $e^{+}e^{-}\to\eta'\pi^{+}\pi^{-}$ by BESIII ($m_{\phi(2170)}=(2108\pm46\pm25)$~MeV/$c^{2}$, $\Gamma_{\phi(2170)}=(138\pm36\pm30)$~MeV)~\cite{Bo} within
two standard deviation.


\section*{\boldmath ACKNOWLEDGEMENTS}
The BESIII collaboration thanks the staff of BEPCII and the IHEP computing center for their strong support. This work is supported in part by National Key R\&D Program of China under Contracts Nos. 2020YFA0406400, 2020YFA0406300; National Natural Science Foundation of China (NSFC) under Contracts Nos. 11625523, 11635010, 11735014, 11822506, 11835012, 11935015, 11935016, 11935018, 11961141012, 12005219, 12022510, 12025502, 12035009, 12035013, 12061131003, 11705192, 11875115, 11875262, 11950410506, 12061131003; the Chinese Academy of Sciences (CAS) Large-Scale Scientific Facility Program; Joint Large-Scale Scientific Facility Funds of the NSFC and CAS under Contracts Nos. U1732263, U1832207, U1832103, U2032111; CAS Key Research Program of Frontier Sciences under Contract No. QYZDJ-SSW-SLH040; 100 Talents Program of CAS; INPAC and Shanghai Key Laboratory for Particle Physics and Cosmology; ERC under Contract No. 758462; European Union Horizon 2020 research and innovation programme under Contract No. Marie Sklodowska-Curie grant agreement No 894790; German Research Foundation DFG under Contracts Nos. 443159800, Collaborative Research Center CRC 1044, FOR 2359, FOR 2359, GRK 214; Istituto Nazionale di Fisica Nucleare, Italy; Ministry of Development of Turkey under Contract No. DPT2006K-120470; National Science and Technology fund; Olle Engkvist Foundation under Contract No. 200-0605; STFC (United Kingdom); The Knut and Alice Wallenberg Foundation (Sweden) under Contract No. 2016.0157; The Royal Society, UK under Contracts Nos. DH140054, DH160214; The Swedish Research Council; U. S. Department of Energy under Contracts Nos. DE-FG02-05ER41374, DE-SC-0012069.


\end{document}